\documentstyle[12pt,psfig,obscite]{obsmag}
\title{The Discovery of a Nearby M Dwarf}

\author{O. Shemmer \\and\\
S. Kaspi \\ School of Physics and Astronomy and the Wise Observatory, \\ The
Raymond and Beverly Sackler Faculty of Exact Sciences, \\ Tel-Aviv
University, Tel-Aviv 69978, Israel}

\begin{document}

\maketitle

\begin{abstract}
We report the discovery of a nearby M dwarf star, found accidentally
while observing the old nova DN Gem at the Wise Observatory. The star
is designated 1200-05296925 in the PMM USNO-A1.0 catalogue and its
coordinates, calculated for 1997 November 27 are: RA=6 55 05.13 Dec = +32 09 54.1 (Equinox J2000, Epoch J1997.90). Astrometric measurements for the star yielded a yearly
proper motion rate of 0.155$\pm$0.002 arcseconds in right ascension and
negligible yearly proper motion rate in declination. The apparent V
magnitude of the star was measured as m$_V$=13.87$\pm$0.16 mag and
spectral identification yielded a spectral type of M3.5Ve$\pm$0.5
subclasses. Using relations between spectral type and absolute V
magnitude in M dwarfs, we arrive at an absolute magnitude of
M$_V$=12.3$^{+1.2}_{-1.1}$ mag, which corresponds to a distance of
21$^{+15}_{-10}$ pc.
\end{abstract}

\section{Introduction}

\begin{figure}
\psfig{figure=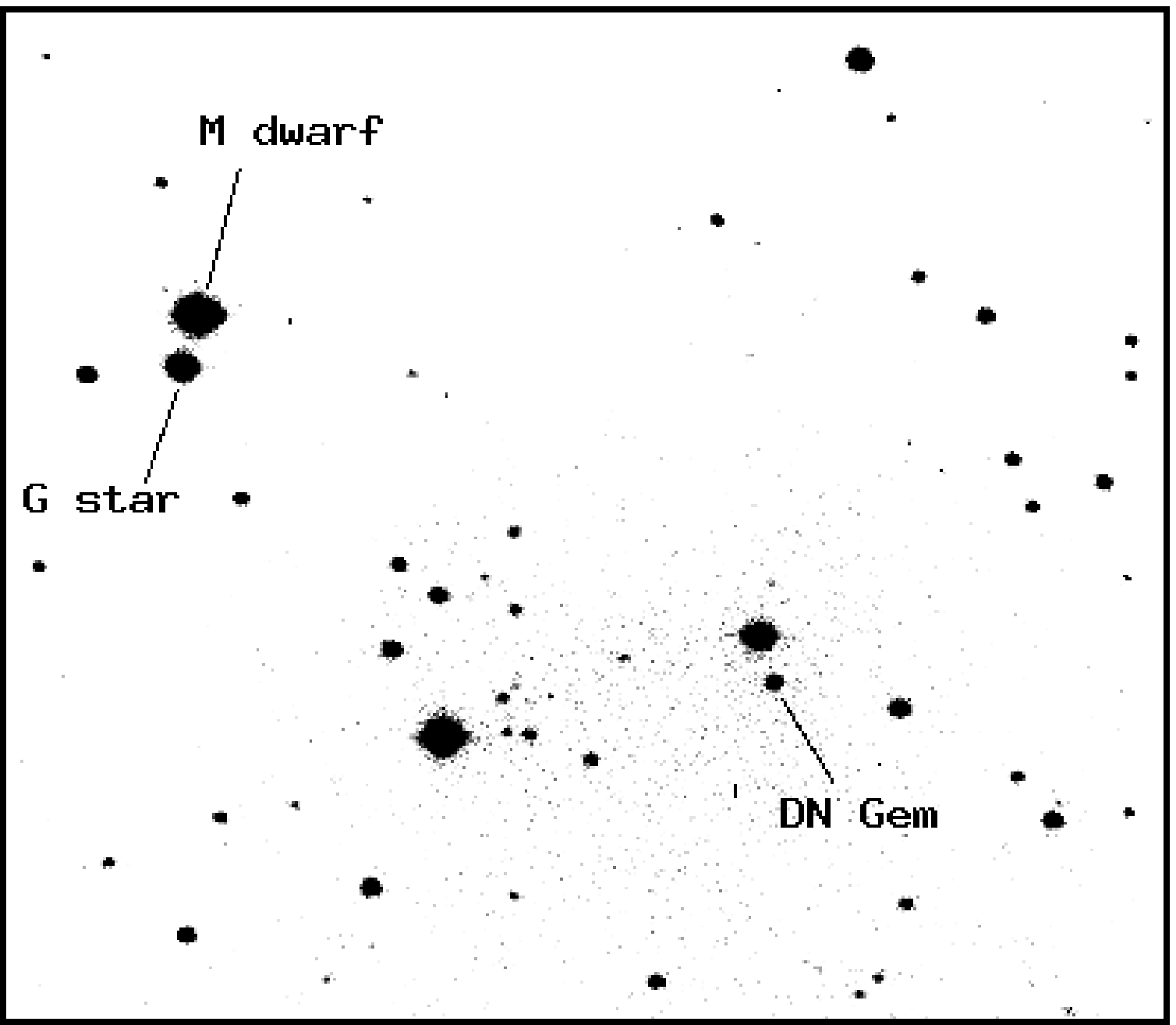} 
\caption{Finding chart of the M dwarf. North is up and East is left in this 4.1$X$3.9 arcminutes field.}
\protect\label{findchart}
\end{figure}

In the course of monitoring periodic light variations in the old nova
DN Gem (nova Gem 1912) at the Wise Observatory\cite{ret98}, on 1997 April 21, one of the stars (hereafter M dwarf)
in the nova's field seemed to have moved a few arcseconds to the west
compared with the nova's old finding chart. Fig.~\ref{findchart} is
part of an image taken on 1997 April 21 to serve as a finding chart.
This visual finding was further checked against other images of the
nova from previous years taken at the Wise Observatory and a Palomar
sky survey plate of the field from 1955. A thorough examination in high
proper motions and nearby stars catalogues, including a search through
the SIMBAD database yielded no information on this star.

Further examination of DN Gem's field, resulted in yet another puzzling
fact in that the Guide Star Catalogue (hereafter GSC) refers to only
one bright object near the position of the M dwarf, whereas there were
in fact two bright stars in the images around that position (separated
by only a few arcseconds). Checking upon the USNO-A1.0 catalogue
resulted also in the appearance of one star at that position, but with
the coordinates of the M dwarf. This puzzle was solved by recognizing
that the M dwarf was too faint in V magnitude to appear in the GSC
catalogue, but appears in the USNO-A1.0 catalogue, and the star about
10 arcseconds due south of the M dwarf appears in the GSC catalogue and not
in the USNO-A1.0.  The fact that both stars have proper motions above
the proper motion standard deviation of the background stars (see
below) suggested a possible binary system. However, spectral
classification carried out later on, showed that it is highly unlikely
that the two stars are physically connected, since the primary object
is an M dwarf and the "companion" to its south is a G star (probably a
mid type G dwarf, hereafter G star), not entirely classified, but
still, having calculated its apparent magnitude,
should lie at least several hundreds of parsecs away. We conclude that
it is a mere coincidence that both stars, quite near angularly to each
other, have proper motions higher than that of the background stars.


\section{Observations and Reductions}

The observations were made with the 40" telescope at the Tel-Aviv
University Wise Observatory between 1995 and 1997. Photometry and
spectroscopy were carried out using the CCD camera and the Faint Object
Spectroscopic Camera (FOSC), respectively. Both instruments use a Tektronics
1024$\times$1024 pixel back-illuminated CCD as a detector\cite{k94}.

During the spectroscopic observations the spectrograph 10 arcseconds -wide long-slit was aligned at PA=163.3 deg in order to simultaneously include
both the M dwarf and the G star in the slit. This way a maximum separation
between the two objects spectra is achieved, minimizing the possibility
of light from one object contaminating the other.  The observation on
1997 November 27 was devoted to photometric calibration of the star's
apparent magnitude using Landolt's standard star system\cite{lan92}.

Reduction was carried out in the standard way using the NOAO
IRAF\footnote{{IRAF (Image Reduction and Analysis Facility) is
distributed by the National Optical Astronomy Observatories, which are
operated by AURA, Inc., under cooperative agreement with the National
Science Foundation.}} {\sc {daophot}} package for the photometry and the
{\sc {specred}} package for the spectroscopy. Since no spectroscopic standards
were observed when the spectra of the stars were taken, the spectrum
was flux calibrated using the Wise Observatory standard coefficients,
which generally do not change significantly from night to night. The
spectra extend from 3980 to 7820\AA \ with a dispersion of 3.8\AA \ per
pixel (about 8\AA \ resolution).

\section{Astrometry}

The exact position and proper motion of the M dwarf was calculated
using the IRAF {\sc {imcoords}} package.  We chose 7 bright stellar candidates
in the field of the M dwarf that had their coordinates listed in the
PMM USNO-A1.0 catalogue and had no known proper motions.  These
coordinates were used to calculate the plate solution for each image of
the observations and for a reference image
extracted from the Palomar Digitised Sky Survey (DSS). Next, the
coordinates of the M dwarf and the G star were calculated using the
plate solution for each image.  The M dwarf coordinates calculated for
the image of 1997 November 27 were RA=6 55 05.13 Dec = +32 09 54.1 (Equinox J2000, Epoch J1997.90).

The position shifts (hereafter residuals) for each star were then
calculated as the (2000) RA \& DEC coordinates difference between each
image and the DSS image. The
standard deviation of the residuals of the 7 background stars is
0.1 arcseconds, which sets a limit to our angular resolution. Results of
the astrometric calculations showed that the 7 background stars have
negligible residuals in both RA and DEC while the M dwarf has a
comparatively large residual in RA, which is 6.6$\pm$0.1 arcseconds and
negligible residual in DEC. The time span between the images of
1955 February 23 (DSS) and 1997 November 27 is 42.76 years, therefore
yielding for the M dwarf a yearly proper motion rate of
0.155$\pm$0.002 arcseconds in RA.  This comparatively large proper
motion might hint on the proximity of the star. The G star on the other
hand, has a proper motion rate in DEC one order of magnitude smaller
than the M dwarf proper motion rate in RA, but still about 5$\sigma$
above the mean residual of the background stars.

\section{Photometry}

On the way to verifying the proximity of the M dwarf, we calibrated the
star's magnitude and made corrections for atmospheric extinction. On
1997 November 27 we took exposures of the star with four different
filters: B,V,R,I (Johnson) and several fields of Landolt's standard
star system before and after the M dwarf exposures, in different values
of the airmass. The standard star fields were chosen to contain several
stars with a variety of $B-V$ \ colours.

Table~\ref{tbl-1} lists the resultant apparent magnitudes in different
filters. Magnitude values are given within an error of 0.16 mag.  By
inspection of Table~\ref{tbl-1} it is evident that the resultant
apparent magnitude of the G star makes it impossible to be nearby,
since even late G dwarfs have absolute V magnitudes of the order of 5
mag.

\section{Spectroscopy}

\begin{figure}
\psfig{figure=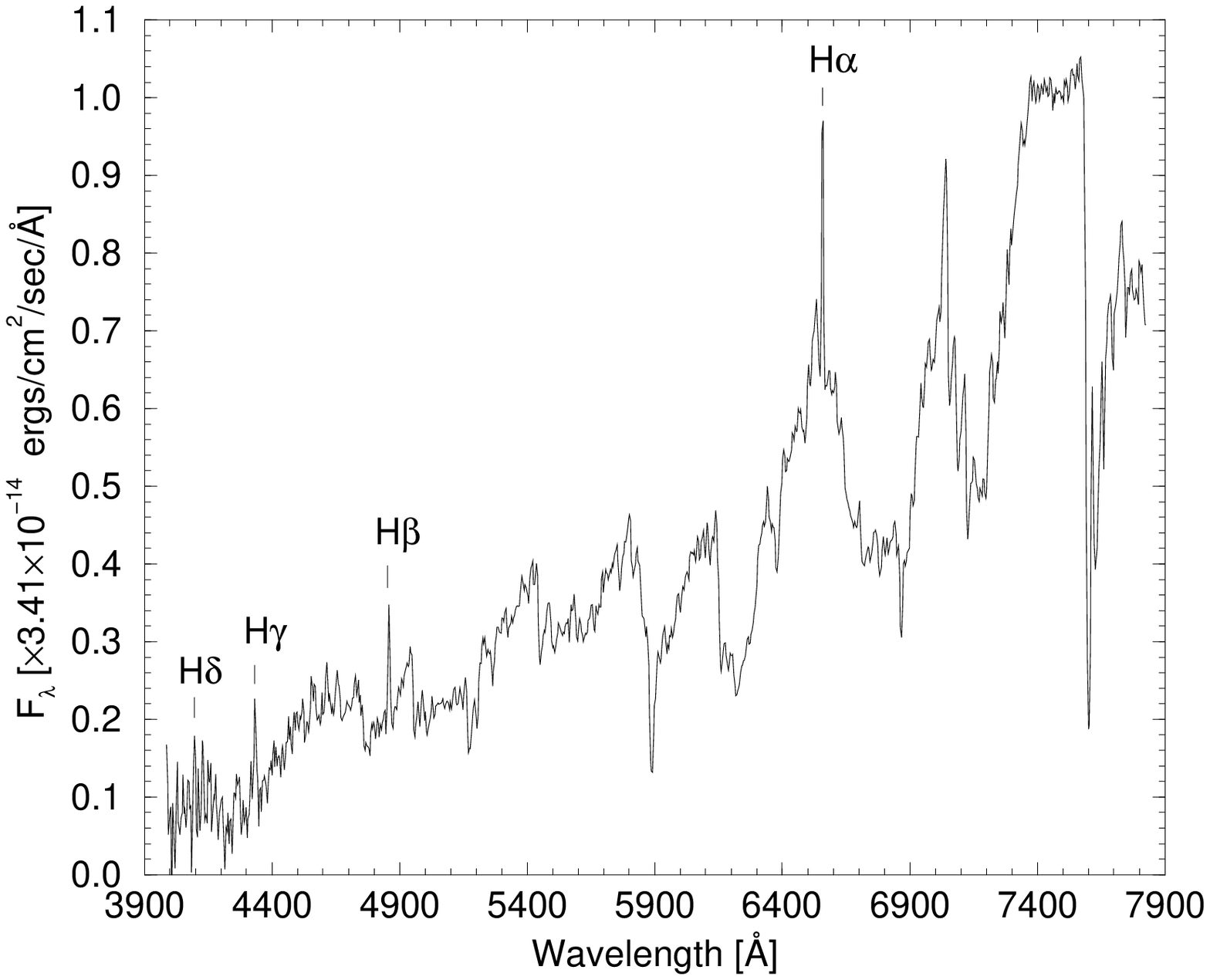} 
\caption{Spectrum of USNO-A1.0 1200-05296925 identified as an M3.5Ve
star. The flux was normalised, according to Kirkpatrick et al. (1991) to
the value at the $\lambda$7500\AA \ continuum level, which was
3.41$X10^{-14}$ ergs/cm$^2$/sec/\AA.}
\protect\label{MDspec}
\end{figure}

The key to our conclusion on the proximity of the M dwarf, lies mainly
on our spectral identification of the star. By preliminary examination
of the spectra we classified the star as an M dwarf and the star to its
south as a G star. In the M dwarf spectra (see Fig.~\ref{MDspec})
several features enabled us to make a further sub$-$classification
using the spectral methods given by Kirkpatrick et al. (1991)\cite{kir91}, and by Henry et al. (1994)\cite{hen94}.  Among the various
features common to M dwarfs, the star's spectrum also shows emission
lines, mainly the Hydrogen Balmer series H$\alpha$, H$\beta$, H$\gamma$
and H$\delta$.  According to Joy \& Abt (1974)\cite{joy74} the proportion of stars
having emission in the Balmer lines grows with spectral type, in
particular for types M3V $-$ M3.5V the ratio of emission line stars
among the M dwarfs is 37.5$\%$.

Sub$-$classification of the M dwarf relied basically on the inspection
of existing spectral templates but also on the apparent spectral
features to specify the spectral type among the dwarfs. Since our
spectrum extends no further to the red than 7820\AA, we were not able
to specify the spectral type in a more precise manner, since many
important features appear in M dwarfs at around 8000\AA. In particular
we only managed to use 3 out of the 8 ratios cited by Kirkpatrick et al. (1991)\cite{kir91} to
make a cross-identification of the spectral type more precisely. Ratio
A, which is the flux ratio between the continuum in the 7020$-$7050\AA
\ band and the CaH band 6960$-$6990\AA \ was calculated to be of the
order of 1.22, suggesting a subclass of M3. Ratio B, which is the flux
ratio between the continuum in the 7375$-$7385\AA \ band and the
Ti{\sc i} band 7353$-$7363\AA \ was calculated to be of the order of
1.07, suggesting a subclass of M4. The ratio of ratio B to ratio A
therefore turns out to be 0.877, suggesting a subclass of M3. It should
be noted that ratio B is almost constant between M3 and M5 and the B to
A ratio is very sensitive to spectral type. It is therefore hard to
identify the spectral type using these 3 ratios alone, even though they
set a low limiting value of about M2 and a high limit on about M5.
However, spectral identification of the discovered M dwarf received
strong support both from the spectral templates with the selected
spectral features and from the observed colour of $B-V$=1.23. The fact
that the VO bands are very weak or absent in our spectrum ensures us
that the star is not considered a late type M dwarf, and the weakness
of the CaH $\lambda\lambda$6382,6389\AA \ bands for instance, or the
weakness of the TiO bands around $\lambda$6600\AA \ prevents it from
being a mid to late M dwarf.  The weakness or total absence of the
Ba{\sc ii} line blended with Ti, Fe and Ca bands at $\lambda$6497\AA,
makes it more likely an early to mid type M dwarf. Even though the
uncertainty in the spectral class still prevails, we establish the
spectral type as M3.5, while it was noted in Henry et al. (1994)\cite{hen94} that spectral
classes of the M dwarfs are in any way uncertain within $\pm$0.5
subclasses.

\begin{figure}
\psfig{figure=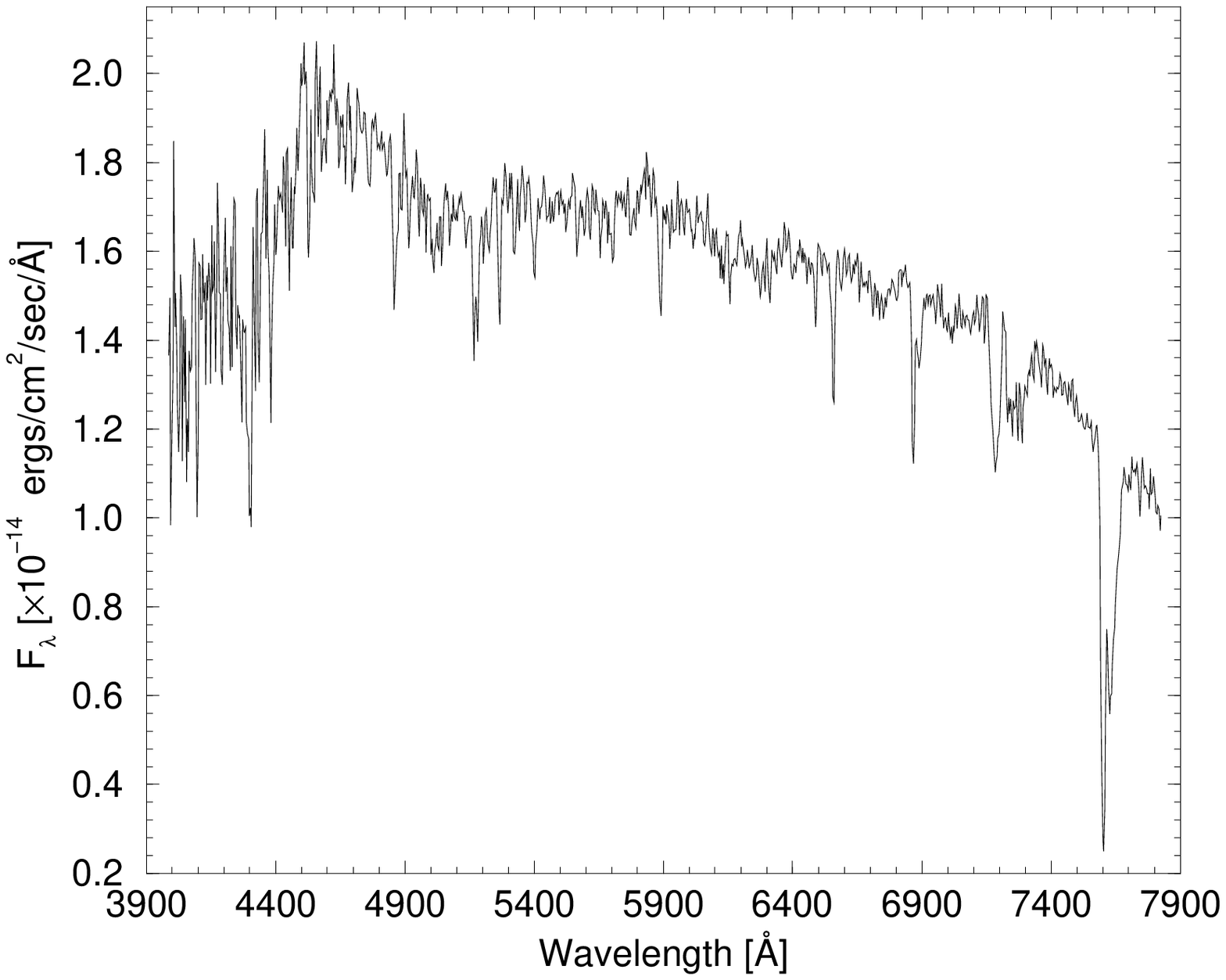} 
\caption{Spectrum of the G star.}
\protect\label{Gspec}
\end{figure}

When analysing the spectrum of the G star we faced more difficulty,
since the main spectral features didn't agree very well with any of the
subclasses or with the luminosity classes, see Fig.~\ref{Gspec}.
Even though, we estimate that the G star is probably a G dwarf, and due
to its faintness must lie at least several hundreds of parsecs away.

\section{Summary}

We have discovered an M dwarf with a comparatively high rate of proper
motion. In order to estimate its proximity, we have used the relation in Henry et al. (1994)\cite{hen94}  between the spectral type of M dwarfs and their
absolute V magnitude, where ST is the spectral type:
\begin{equation}
\label{equation 1}
M_V=0.101(ST)^2+0.596(ST)+8.96 \  .
\end{equation}
The absolute V magnitude obtained by equation~\ref{equation 1} is
within an error of 0.5 mag. Therefore, for an M3.5$\pm$0.5 dwarf, we
arrive at an absolute V magnitude of M$_V$=12.3$^{+1.2}_{-1.1}$ mag.
Using the apparent V magnitude of the M dwarf given in Table~\ref{tbl-1} we can
finally calculate the distance modulus of the M dwarf to be
$\mu=1.6^{+1.2}_{-1.4}$ which determines a distance of 21$^{+15}_{-10}$
pc (neglecting galactic extinction).  We conclude that the M dwarf is
nearby although it does not appear in the catalogue of nearby stars\cite{gli91}. Future astrometry, in particular trigonometric
parallax measurements, which are too sensitive for the Wise Observatory
equipment, may yield a more precise value of the distance to the star.

\section{Acknowledgments}

We would like to thank Tsevi Mazeh for guiding us through the issue of
nearby stars and M dwarfs. We would also like to thank Dan Maoz for
helping us with his standard star reduction program, Liliana Formiggini
for her assistance in spectral classification and Elia M. Leibowitz and
Alon Retter for their helpful remarks. This research has made use of
the SIMBAD database, operated at CDS, France. Astronomy at the Wise
Observatory is supported by grants from the Israeli Academy of
Sciences.

\bibliographystyle{obsmag}
\bibliography{iau_journals,MdwarfV2310}

\begin{table}[p]
	\caption{Apparent Magnitudes and Colours. \label{tbl-1}}
	\begin{tabular}{lcccccc}
Filter/colour & B & V & R & I & $B-V$ & $V-I$ \\ 
M dwarf & 15.10 & 13.87 & 12.64 & 11.03 & 1.23 & 2.84 \\
G star & 13.87 & 13.35 & 12.83 & 12.35 & 0.52 & 1.00 
	\end{tabular}
\end{table}

\end{document}